\documentclass[runningheads]{llncs}
\usepackage[T1]{fontenc}
\usepackage{graphicx}
\usepackage[outdir=./figures/]{epstopdf}

\usepackage{cite}
\usepackage{amsmath,amssymb,amsfonts}
\usepackage{algorithm}
\usepackage{algpseudocode} 
\usepackage{graphicx}
\usepackage{textcomp}
\usepackage{comment}
\usepackage{todonotes}
\usepackage{booktabs}
\usepackage{caption}
\usepackage{subcaption}
\usepackage{xcolor}
\usepackage[hidelinks, bookmarks=false]{hyperref} 
\usepackage{mathrsfs}
\usepackage{array}
\def\BibTeX{{\rm B\kern-.05em{\sc i\kern-.025em b}\kern-.08em
		T\kern-.1667em\lower.7ex\hbox{E}\kern-.125emX}}

\usepackage{letltxmacro}
\usepackage{xparse}
\usepackage{multirow}
\usepackage{pgfplots}
\usepackage{tikz}
\usepackage{mathrsfs}

\def\equationautorefname~#1\null{Eq.(#1)\null}

\newcommand{\blue}[1]{\textcolor{blue}{\textbf{#1}}}
\begin{document}
\title{Lossless preprocessing of floating point data to enhance compression\thanks{This work was supported by the IoTalentum Project within the Framework of Marie Skłodowska-Curie Actions Innovative Training Networks (ITN)-European Training Networks (ETN), which is funded by the European Union Horizon 2020 Research and Innovation Program under Grant 953442.}}
%
%
\author{Francesco Taurone
\and
Daniel E. Lucani
\and
Marcell Fehér
\and
Qi Zhang
}
\authorrunning{F. Taurone et al.}
%
\institute{DIGIT, Department of Electrical and Computer Engineering, Aarhus University
\email{\{francesco.taurone,daniel.lucani,sw0rdf1shqz,qz\}@ece.au.dk}}
\maketitle              
\begin{abstract}
	Data compression algorithms typically rely on identifying repeated sequences of symbols from the original data to provide a compact representation of the same information, while maintaining the ability to recover the original data from the compressed sequence.
	Using data transformations prior to the compression process has the potential to enhance the compression capabilities, being lossless as long as the transformation is invertible.
	Floating point data presents unique challenges to generate invertible transformations with high compression potential.
	This paper identifies key conditions for basic operations of floating point data that guarantee lossless transformations. Then, we show four methods that make use of these observations to deliver lossless compression of real datasets, where we improve compression rates up to 40~\%.
\end{abstract}

\section{Introduction}

Data management is a key aspect of most IoT related applications, since data generated by sensors usually has to be transmitted, stored and analyzed. A common practice to store time-series datasets is to compress them before archival, so to limit the needed database space. Before compression, data is usually preprocessed, undergoing a transformation process where the content of the dataset is adjusted for compression, which can potentially enhance the compressibility of the dataset.
There are various preprocessing approaches that the user could adopt, ranging from rounding the decimals, to changing domain of the signal with the discrete wavelet transform \cite{dct}, or dropping bits that carry minimal information  \cite{infoPreprocessing}. One key distinction among preprocessing transforms is whether they are lossy or lossless. Given a transformation function $f$ and a dataset $\mathrm{DS}$, we say that $f$ is \textit{lossless} when $f^{-1}(f(\mathrm{DS})) = \mathrm{DS}$, \textit{lossy} otherwise.
The goal of this paper is to characterize lossless operation on floating point data and to propose 4 lossless preprocessing techniques enhancing the compressibility for mono-dimensional floating point datasets, meaning that the preprocessed dataset $f(\mathrm{DS})$, once compressed, is smaller in size than the compressed vanilla $\mathrm{DS}$. We quantify compression in terms of compression ratio ($\mathrm{CR}$), defined as
\begin{equation}
	\mathrm{CR} = \frac{\text{Compressed dataset size in bits} + \text{Compression metadata}}{\text{Uncompressed dataset size in bits}}
	\label{eq:CR}
\end{equation}
where the metadata might be needed to be able to undo the compression.
Since these transformations operate on each individual floating point number, they are naturally compatible with compression methods with fine-grained random access capabilities, such as Generalized Deduplication \cite{GD_RandomAccess}, which are useful for computing analytics directly on the compressed data \cite{Glean}.

\subsection{Compressing floating point numbers}
\label{sec:floatingPoint}
The most common standard for floating point numbers is the IEEE-754\cite{IEEE754}, where a number $x$ is represented by three components: the sign ($\mathrm{S}$), the exponent ($\mathrm{E}$) and the mantissa ($\mathrm{M}$). In this paper, we assume the use of double precision, namely 64 bits per number. When interpreted as unsigned integers, $x$ is
\begin{equation}
	x = (-1)^S \cdot 2^{E-B}\cdot(1 + M\cdot2^{-l}),
	\label{eq:floatingpoint}
\end{equation}
where $B = 1023$ is the bias of the exponent, and $l = 52$ is the length of the mantissa in bits. Out of the 64 bits, $1$ is the sign bit, $11$ are the exponent and the remaining $52$ bits are the mantissa. Each mantissa bit is $m_i$, with $i\in\left\{1, 2, \dots, 52\right\}$ and $m_1$ being the most significant one. From \autoref{eq:floatingpoint}, we see that all $x$ such that $\left|x\right| \in [2^{E^*}, 2^{E^*+1})$ have the same exponent $E^* + B$.

As studied in \cite{additionmethod}, a possible idea to improve compressibility is to maximize the number of bits shared by all numbers in the collection we want to compress. We say that the i-th bit is \textit{shared} when all numbers int eh dataset have the same value for the i-th bit, meaning that if all bits are shared, the dataset is composed of a series of equal numbers. The work in \cite{additionmethod} empirically shows that having more shared bits generally results in better compression, with both standard compressors, like zlib \cite{zlib}, and ones based on deduplication, used in \autoref{sec:results}. However, the preprocessing technique $\bar{f}$ showed in \cite{additionmethod} is lossy: in the following, we elaborate on it to propose lossless preprocessing techniques.
\section{Key observations}

\subsection{Lossless operations on floating point numbers}
\label{losslessOperations}
Numbers represented with a finite amount of bits will necessarily have finite precision. IEEE-754 can represent a subset of all numbers on the real line, while values that are not in this subset get approximated to one that is. In this context, we refer to the precision of a number $x = (-1)^{S_x} \cdot 2^{E_x-1023}\cdot(1 + M_x\cdot2^{-52})$ as the distance between $x$ and the next representable floating point number. We call this quantity \textit{unit in the last place}, or $\mathrm{ULP}(x)$, calculated as
\begin{equation}
	\mathrm{ULP}(x) = 2^{E_x - 1023 - 52} 
\end{equation}
The reason why manipulating floating point numbers can result in losses is that $\mathrm{ULP}(x)$ is not a constant in floating point, but rather a function of the number's unbiased exponent. 
Transformations whose output results in $\mathrm{ULP}(x)$ different from the original could potentially lead to information losses. In order to explain this phenomenon, we introduce the operations $\oplus$, $\otimes$, $\ominus$, $\oslash$, which represent addition, multiplication, subtraction and division according to IEEE-754. For our purposes, we can think of them as being the theoretical operations, followed by rounding to the nearest representable number. 

An example of potential losses due to floating point precision can be shown by considering $x = 3.5$, the transform $f(x) = x \oplus 10^{16}$ with its inverse $g(x) = x \ominus 10^{16}$ and their theoretical counterpart $f^\prime(x) = x + 10^{16}$ with its inverse $g^\prime(x) = x - 10^{16}$. While $g^\prime(f^\prime(x)) = x \, ,\forall x\in\mathbb{R}$, the same is not true in floating point, since $g(f(3.5)) = 4.0 \neq 3.5$. 

While there are multiple strategies to limit the effects of this approximation error, as described in Section 5.3 of  \cite{FPHandbook}, this paper aims at devising strategies to avoid it completely. In \autoref{losslessAddition}, we discuss particular scenarios for which summation does not produce error, and in \autoref{losslessMultiplication} we analyze multiplication.

\subsubsection{Lossless addition}
\label{losslessAddition}
In the scenario where $x \in [2^{E^*}, 2^{E^*+1})$ and $f(x) = A \oplus x$ is such that $f(x) \in [2^{E^*+1}, 2^{E^*+2})$ and $A \in [2^{E^*}, 2^{E^*+1})$, the operation is lossless as long as both $A$ and $x$ have the same least significant mantissa bit $m_{52}$. The reason is that during an addition under these conditions, a bit called \textit{guard bit} has to be rounded away, since $\mathrm{ULP}(\cdot)$ goes from $\mathrm{ULP}(x) = 2^{E^*-52}$ to   $\mathrm{ULP}(f(x)) = 2^{E^*-51}$. However, if both addends have the same $m_{52}$, the guard bit is guaranteed to be equal to zero, requiring no rounding. This result is summarized in \autoref{tab:losslessAddition} and more details on how addition works in floating point in Section 7.3 of \cite{FPHandbook}.
\begin{table}[h]
	\centering
	\vspace{-0.5cm}
	\caption{Mantissa least significant bits for lossless addition $y = x \oplus A$. The blue results indicate a lossless transformation, namely $y\ominus A = x$.}
	\begin{tabular}{lr|cccc}
		\toprule
		$m^y_{52}$&& \multicolumn{4}{c}{$m^x_{51} m^x_{52}$}\\
		&& $00$ & $01$ & $10$ & $11$ \\
		\hline
		\parbox[t]{2mm}{\multirow{4}{*}{\rotatebox[origin=c]{90}{$m^A_{51} m^A_{52}$}}}&$00$&\blue{0}&0&\blue{1}&0\\
		&$01$&0&\blue{1 }&0&\blue{0 } \\
		&$10$&\blue{1}&0&\blue{0}&0 \\
		&$11$&0&\blue{0}&0&\blue{1}\\
		\bottomrule
	\end{tabular}
	\label{tab:losslessAddition}
	\vspace{-0.3cm}
\end{table}

Another particular useful scenario is the addition $y = x \oplus A$ where $x, y \in [2^{E^*}, 2^{E^*+1})$ and $A \in [2^{\tilde{E}}, 2^{\tilde{E}+1})< 2^{E^*}$. It can be shown that given the floating point addition process detailed in \cite{FProunding}, this operation is lossless when
\begin{equation}
	m_i = 0 \quad\text{for } i \in \left\{52 - \left(E^* - {\tilde{E}} +1\right), \dots,  52\right\}.
	\label{eq:losslessAddWithinRegion}
\end{equation}

\subsubsection{Lossless multiplication}
\label{losslessMultiplication}
In the scenario where $x \in [2^{E^*}, 2^{E^*+1})$ and $f(x) = x \otimes M$ is such that $f(x) \in [2^{E^*+1}, 2^{E^*+2})$, the operation is guaranteed to be lossless as long as $M \geq 2$.
\begin{figure}[tb]
	\centering
	\includegraphics[width = 0.7\textwidth]{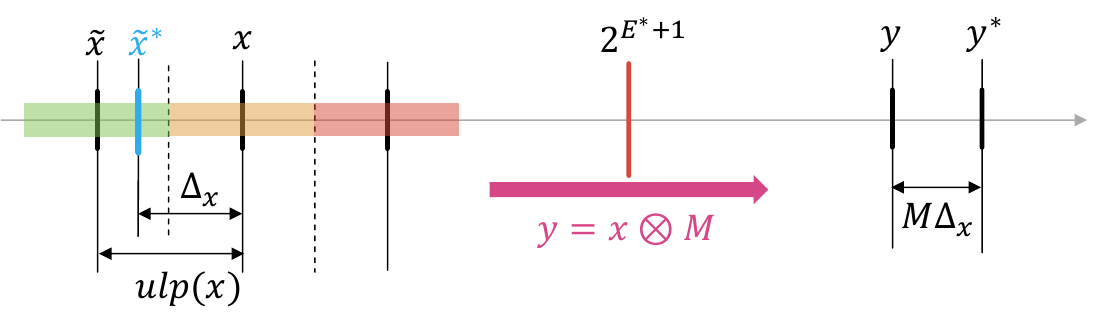} 
	\caption{Representation of the loss with $f(x)= y = x \odot M$ and $f^{-1}(x) = \tilde{x} = y \oslash M$. Assuming the use of the \textit{round to nearest} approach, all real numbers in the same color region get rounded to the same floating point number.}
	\label{fig:losslessMultiplication}
\end{figure}
To prove it, we consider \autoref{fig:losslessMultiplication}, where $y^* = x \cdot M$, $y= x \otimes M$, $\tilde{x}^* = y / M$ and $\tilde{x} = y \oslash M$. In order for this transformation to be lossless, we need
\begin{equation}
	\left|x-\tilde{x}^*\right|=\Delta_x < \mathrm{ULP}(x) / 2
	\label{eq:losslessMultDisequality}
\end{equation}
since this would result in $\tilde{x} = x$ under the\textit{ round to nearest }rounding scheme.  Since $y = \tilde{x}^* \cdot M =  (x + \Delta_x) \cdot M$, we have$y - y^* =  M \cdot \Delta_x$, which combined with \autoref{eq:losslessMultDisequality}, results in the condition for lossless multiplication being
\begin{equation}
	y - y^* < M \cdot \frac{\mathrm{ULP}(x)}{2} .
	\label{eq:losslessMultiplicationConditionForLossless}
\end{equation}
Since $\mathrm{ULP}(y) = 2 \cdot \mathrm{ULP}(x)$ and $y - y^* < \frac{\mathrm{ULP}(y)}{2}$ because of the rounding method, we have $	y-y^*<\mathrm{ULP}(x) <  M \cdot \frac{\mathrm{ULP}(x)}{2}$, fulfilling \autoref{eq:losslessMultiplicationConditionForLossless} when $M \geq 2$.

\section{Lossless dataset manipulations for better compression}
\label{sec:datasetManipulations}
In order to increase the number of shared mantissa bits in the dataset, our goal is to ensure that all numbers after preprocessing lie in the portion of the real line where the mantissa bits are guaranteed to be as we desire. Given a dataset $\mathrm{DS}$, in order to guarantee that the $D$ most significant mantissa bits are shared,
the datapoints $x$ would all have to be such that
\begin{equation}
	\left|x\right|\in\left[2^E, 2^E + 2^{E-D}\right]\quad \text{with } E \in \left[-1022, 1023\right] \quad \forall x \in \mathrm{DS}.
\label{eq:sameMantissaBitsRegions}
\end{equation}
Next, we present four techniques to losslessly place the numbers in the dataset in these preferred regions, supposing without loss of generality and for clarity sake to have a dataset where all numbers have the same exponent. This can be easily generalized by storing as metadata the information on the original exponent of each sample.
We named these techniques \textit{compact bins, multiply and shift, shift and separate even from odd, shift and save evenness.}

\subsection{Compact bins}
\label{binsTechnique}
\begin{figure}[tbh]
	\centering
	\includegraphics[width = 0.8\textwidth]{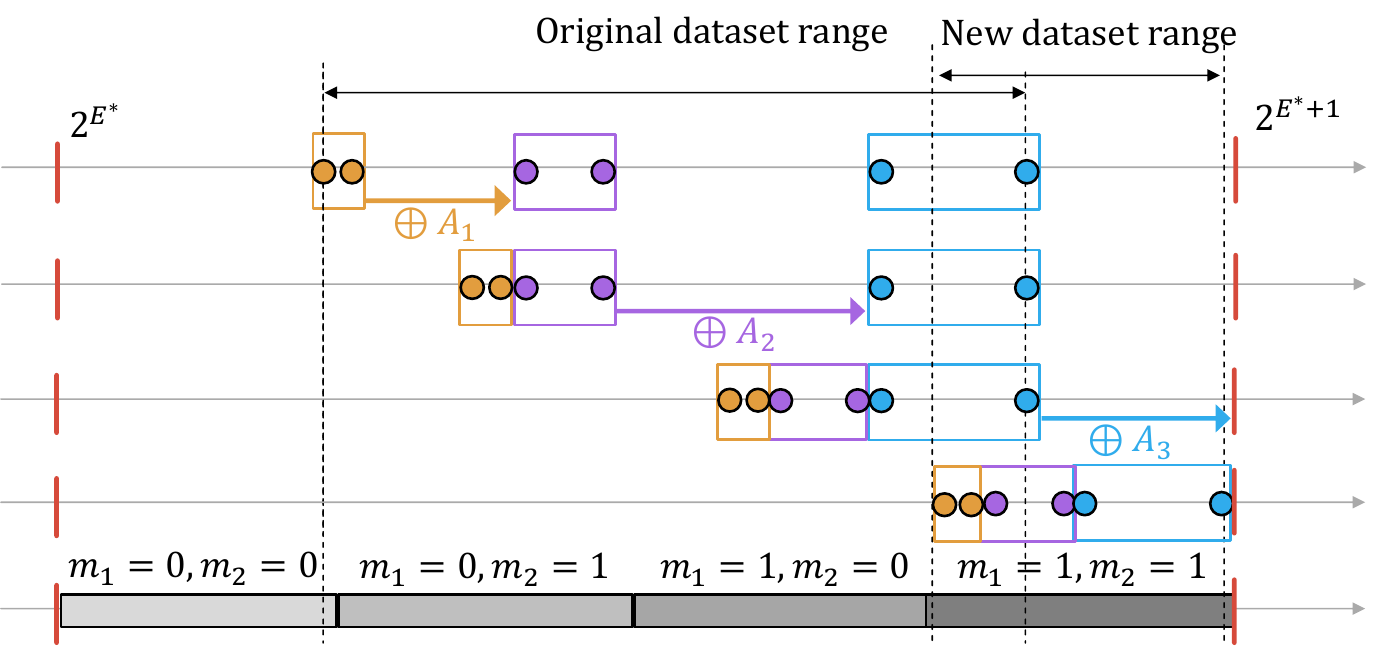} 
	\caption{\textbf{Illustration of method \textit{compact bins}}. In this example, by using 3 bins all numbers in the new dataset are guaranteed to have $m_1 = 1$ and $m_2 = 1$, while there was no such guarantee in the original dataset.}
	\label{fig:bins}
\end{figure}
In the scenario where $x \in \left[2^{E^*}, 2^{E^* +1}\right)$, if we cluster this set of values into bins and shift the bins so they they are closer to each other, we are effectively reducing the range of values of the transformed dataset, possibly making it fit to a region where some mantissa bits are guaranteed to be shared. An example of this binning process is in \autoref{fig:bins}, where we use 3 bins to ensure $m_1 = 1$ and $m_2 = 1$ $\forall x \in {\mathrm{DS}}$. The optimal amount of bins depends on the distribution of values in the real axis. Two conditions are needed in order for this process to be lossless. First, supposing to use $k$ bins on a dataset with $\ell$ unique values, we have to store $(k-1)\cdot \lceil \log_2\left(\ell\right)\rceil$ bits as metadata representing the boundaries of the bins, as well as the $k$ values $\left\{A_1, \dots, A_k\right\}$ used to shift the bins. The size of metadata in bytes is $\mathrm{Z} = (k\cdot 64 + (k-1)\cdot b )/8$. Secondly, the $A_i$ have to be chosen so that the operation $y = x \oplus A_i$ is lossless. We can use \autoref{eq:losslessMultiplicationConditionForLossless} to select the correct $A_i$ values.

\subsection{Multiply and shift}
\label{multiplyTechnique}

As per \autoref{eq:sameMantissaBitsRegions}, there is one portion of the real axis per exponent region guaranteeing that the $D$ most significant bits are equal to 1. Using the lossless operations in \autoref{losslessOperations}, we can manipulate the dataset so that the modified DS has all values lying in those regions.
\begin{figure}[tbh]
	\centering
	\includegraphics[width = 0.9\textwidth]{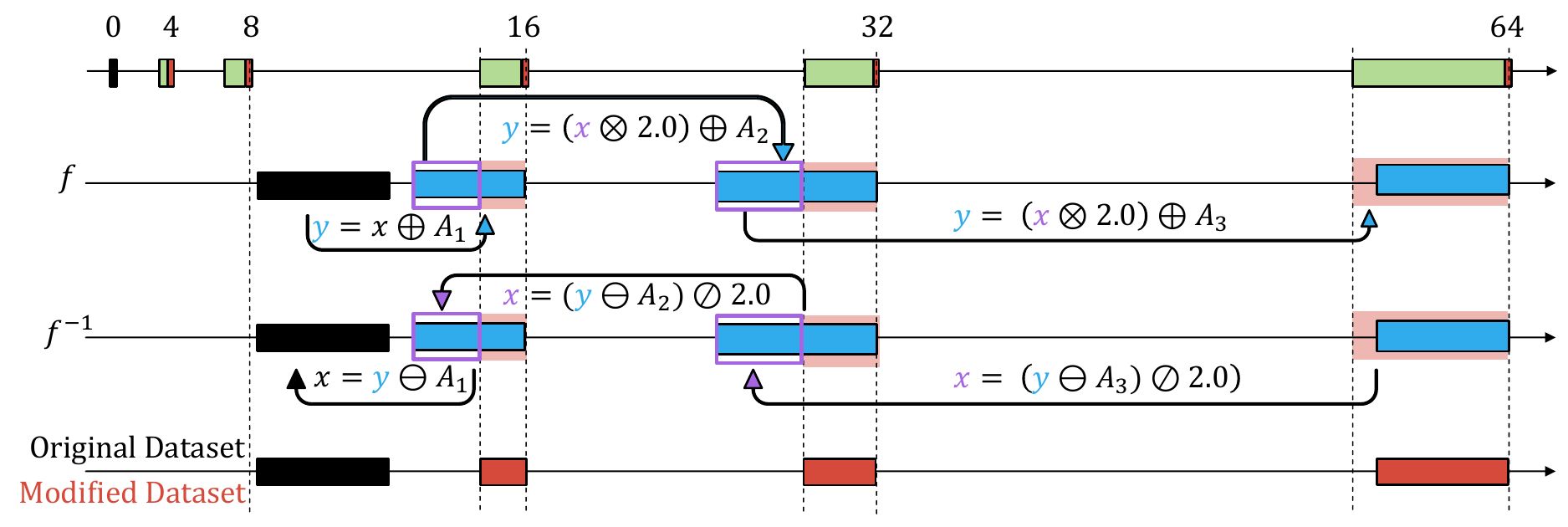} 
	\caption{\textbf{Illustration of method \textit{multiply and shift}}. From the original dataset in black, we iteratively apply \autoref{eq:multAndShiftTransform} to obtain a transformed dataset in red lying on regions in green, where all numbers are guaranteed to share at least the first $D$ mantissa bits. We can recover the original dataset with the inverse transform since all operations are lossless.}
	\label{fig:multAndShiftDiagram}
\end{figure}
The \textit{multiply and shift} transform we propose, represented in \autoref{fig:multAndShiftDiagram}, involves one multiplication and one addition to move to the next exponent region, applied interatively until all modified numbers lie on the portion of the real axis where the most significant $D$ mantissa bits are guaranteed to be shared. Assuming that $x \in \left[2^{E^*}, 2^{E^* +1}\right)$ $\forall x \in \mathrm{DS}$, the operation is 
\begin{equation}
	f(x) = \left(2.0 \otimes x\right) \oplus A, \quad \text{with } \, A = 2^{E^*-D + 1} - 2\cdot\mathrm{ULP}\left(2^{E^*+1}\right),
\label{eq:multAndShiftTransform}
\end{equation}
where $A$ has to be rounded down if necessary to the first value fulfilling \autoref{eq:losslessAddWithinRegion}.
As we can see in \autoref{fig:multAndShiftDiagram}, after an iteration, only the values out of the region with desired common bits have to go through another one, progressively shaving the dataset under analysis. A limitation of this algorithm is that the multiplication by a factor of 2.0 effectively enlarges the window of values of the transformed portion of the dataset, increasing the number of needed iterations. However, $M=2$ is the smallest lossless factor, as per \autoref{eq:losslessMultiplicationConditionForLossless}.
In terms of metadata, we need to store $D$ and $A_1$, since all $A_i$ with $i \neq 1$ can be computed by knowing the exponent of the original dataset and $D$. In order to invert the transformation, given the transformed dataset, we can iteratively apply the inverse transformation $f^{-1}(y) = \left(y \ominus A\right) \oslash 2.0$ on each element belonging to the rightmost exponent region, concluding with the inverted shift stored as metadata.

\subsection{Shift and separate even from odd}
\label{addTechniqueAndSeparate}
In order to alleviate the drawbacks of \textit{multiply and shift} in \autoref{multiplyTechnique}, we can substitute the multiplication with an addition, while carefully selecting the addendum so that it fulfills the conditions in \autoref{losslessAddition}. Specifically, given the operation $y = x \oplus A$ with $x, A \in \left[2^{E^*}, 2^{E^* +1}\right)$, their mantissas $M_x$ and $M_A$ have to have the same evenness, namely $m_{52}^x = m_{52}^A$.
We do that by ensuring that all $y$ resulting from $x$ with even mantissas lie in a portion of the real axis that does not overlap with the one for the odds. In this way, we can distinguish the two during the inverse transformation.
\begin{figure}[tbh]
	\centering
	\includegraphics[width = 0.9\textwidth]{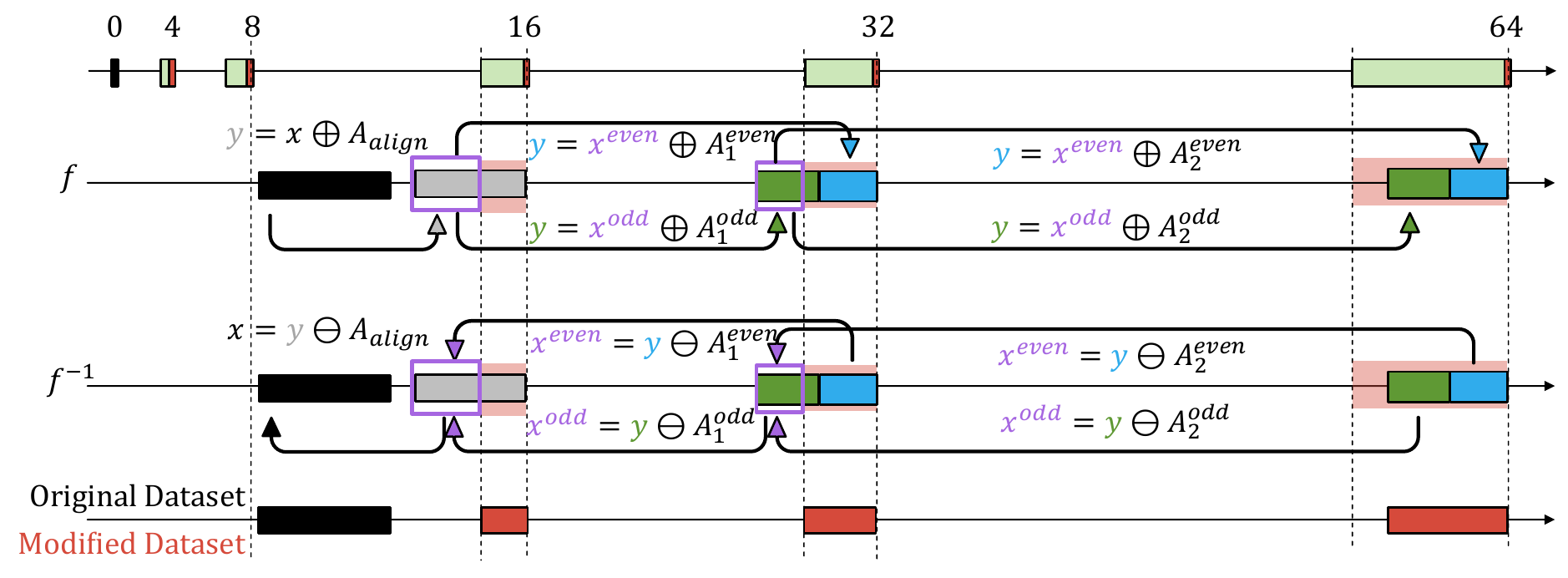} 
	\caption{\textbf{Illustration of method \textit{shift and separate even from odd}}.We iteratively apply \autoref{eq:addTechniqueAndSeparateDirectTransf}, ensuring that the output of $x$ with even and odd mantissas lie in different portions of the real axis ath every iteration.}
	\label{fig:addDifferentPortionsDiagram}
\end{figure}
As depicted in \autoref{fig:addDifferentPortionsDiagram}, the transformation differs depending on the evenness of the mantissas as input, resulting in
\begin{equation}
	f(x)=
	\begin{cases}
		x \oplus A_i^\mathrm{even} & \text{if } \mod(M_x, 2) = 0 \\
		x \oplus A_i^\mathrm{odd} & \text{if } \mod(M_x, 2) = 1 .
	\end{cases}
\label{eq:addTechniqueAndSeparateDirectTransf}
\end{equation}
$A_i^\mathrm{even} $ and $A_i^\mathrm{odd}$ at the i-th iteration are computed so to fulfill \autoref{eq:losslessAddWithinRegion}.
In particular, at th i-th iteration, with $x, A_i^\mathrm{even} , A_i^\mathrm{odd}  \in \left[2^{E^*}, 2^{E^* +1}\right)$ and $D$ desired shared most significant mantissa bits, they are
\begin{equation}
	A_i^\mathrm{even} = 2^{E^*+1} + 2^{E^* - D}, \quad A_i^\mathrm{odd} = A_i^\mathrm{even} - W_i,
\end{equation}
where $W_i$ represents the length of the portion of the dataset to be processed at the i-th iteration. It is calculated starting from the original dataset $\mathrm{DS}$ with $W_{i+1} = 2\cdot W_i - 2^{E^* - D}$, where $W_0 = \max(\mathrm{DS}) - \min(\mathrm{DS})$.
In terms of metadata, we need to store the initial shift $A_{align}$, $D$ and $W_0$, since all $A_i$ can be reconstructed from those.
The i-th step of the inverse transformation of $y \in \left[2^{E^*}, 2^{E^* +1}\right]$ is
\begin{equation}
	f^{-1}(y)=
	\begin{cases}
		y \ominus A_i^\mathrm{even} & \text{if } y > 2^{E^* +1}-W_i \\
		x \ominus A_i^\mathrm{odd} & \text{if } y \leq 2^{E^* +1}-W_i .
	\end{cases}
\label{key}
\end{equation}
With this method we strategically choose $A_i^\mathrm{even}$ and $A_i^\mathrm{odd} $ so that we can guess the evenness of the original $x$ by checking the position of $y$ on the real axis. The drawback is that we need to enlarge the range of values of the transformed dataset at every iteration, similarly to what happened with the multiplication by $2.0$ in \autoref{multiplyTechnique}.

\subsection{Shift and save evenness}
\label{addSaveEvenness}	
In order to limit the scaling up of the transformed dataset at every iteration in \textit{shift and separate even from odd}, we could store the evenness information as a single metadata bit per sample, meaning $n$ bits per iteration for a dataset of $n$ elements. The increased size of metadata is compensated by a drastic reduction in the number of iterations needed to end the procedure.
The direct transformation formulas is the same as per \autoref{eq:addTechniqueAndSeparateDirectTransf}, where the only difference is that the calculation of $A_i^\mathrm{odd}$ becomes $A_i^\mathrm{odd} = A_i^\mathrm{even}$.

\begin{figure}[tbh]
	\centering
	\includegraphics[width = 0.9\textwidth]{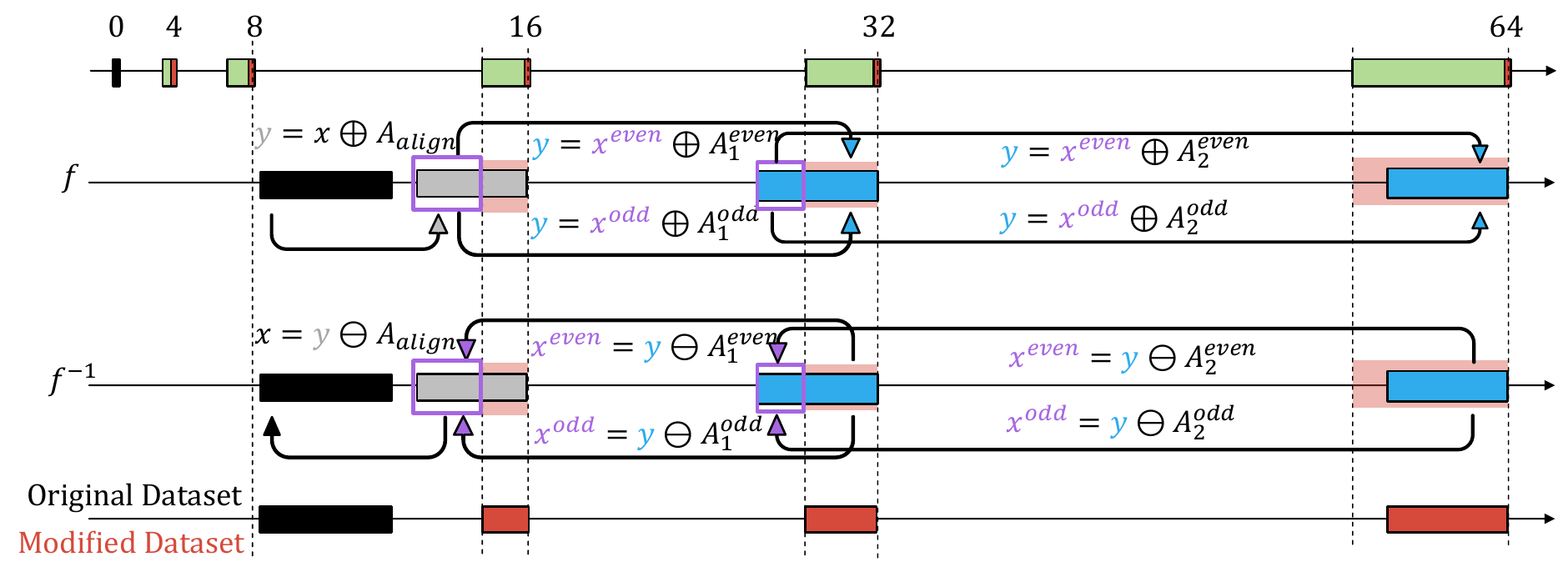} 
	\caption{\textbf{Illustration of method \textit{shift and save evenness}}. While similar in concept to \autoref{fig:addDifferentPortionsDiagram}, we use metadata to distinguish whether to apply the inverse transformation with $A_i^\mathrm{odd}$, namely having odd mantissas, or with $A_i^\mathrm{even}$, which has it even.}
	\label{fig:addSaveEvennessDiagram}
	\vspace{-0.5cm}
\end{figure}

\section{Results}
\label{sec:results}

\begin{figure}[tbh]
	\centering
	\includegraphics[width = 0.9\textwidth]{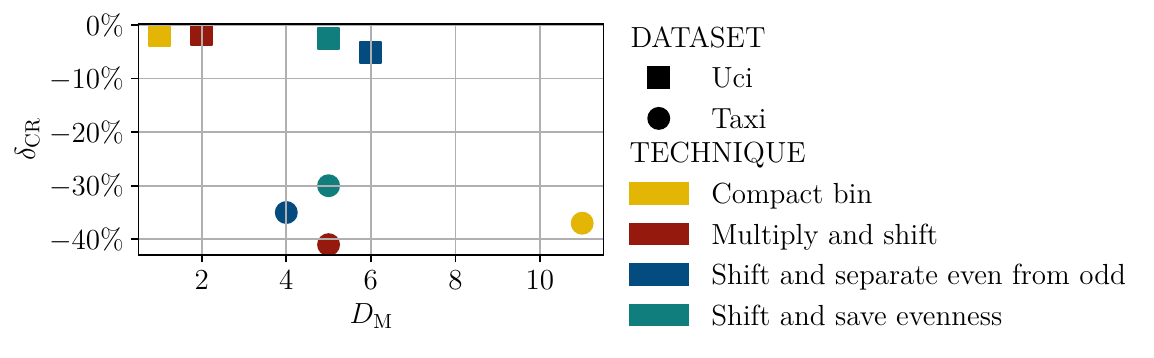} 
	\caption{Comparison of the best results from \autoref{fig:resultTaxi} and \autoref{fig:resultUci}, where lower $\delta_{\mathrm{CR}}$ are better.}
	\label{fig:comparisonOfResults}
\end{figure}

\begin{figure}[t]
	\centering
	\begin{subfigure}[b]{\textwidth}
		\centering
		\includegraphics[width =\textwidth]{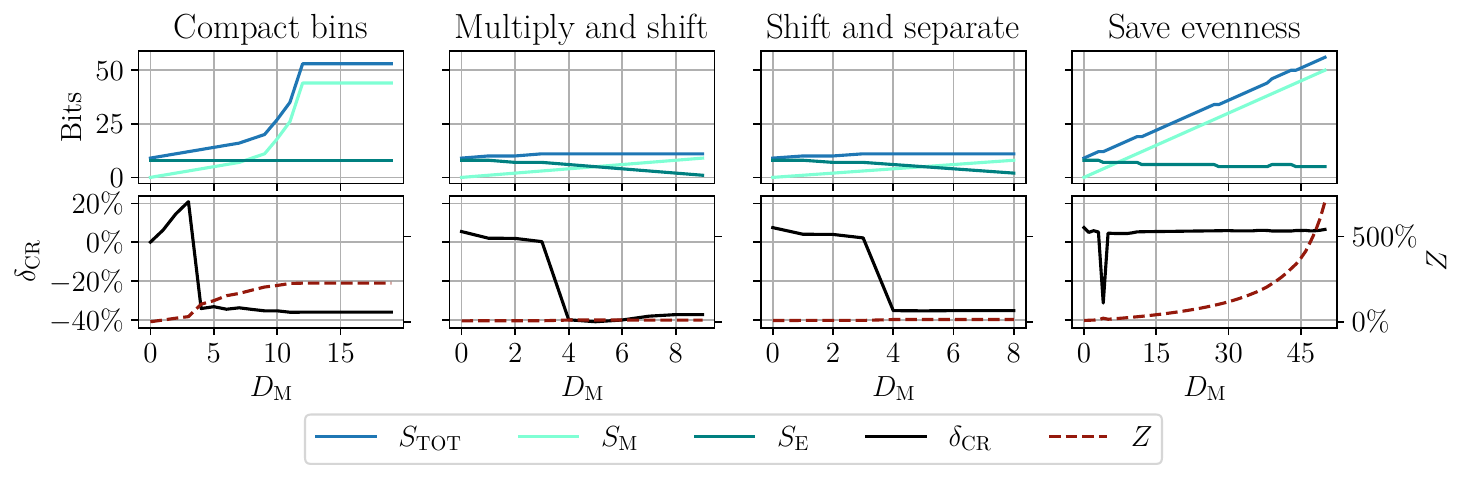} 
		\caption{Chicago-taxi-trips-fares \cite{TaxiDataset}}
		\label{fig:resultTaxi}
	\end{subfigure}
	\hfill
	\begin{subfigure}[b]{\textwidth}
		\centering
		\includegraphics[width =\textwidth]{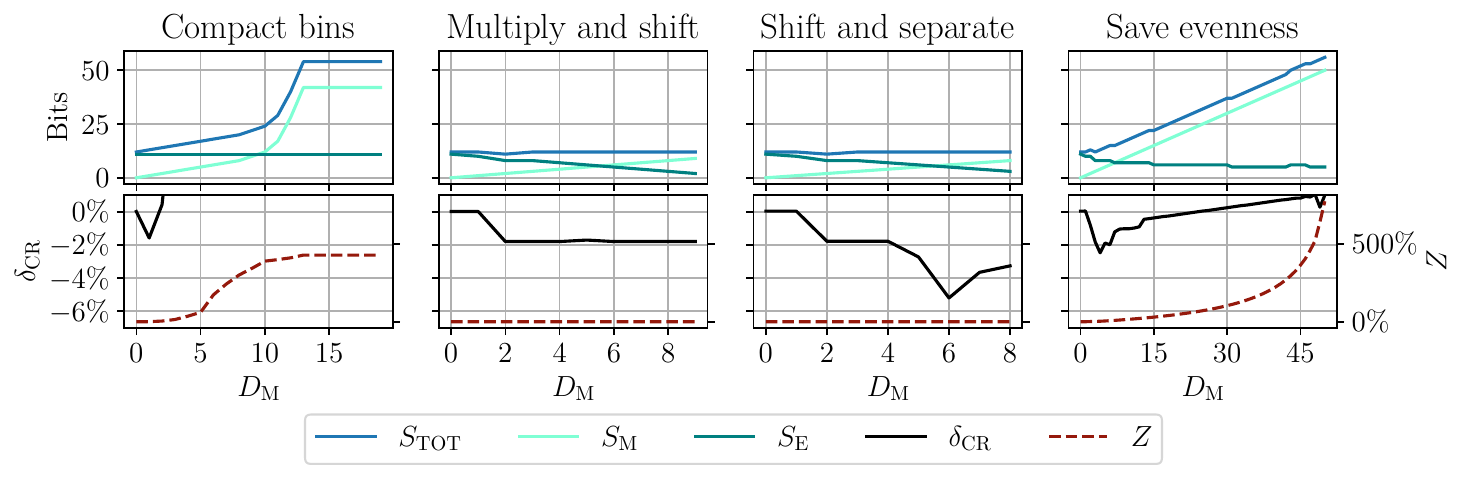} 
		\caption{UCI-gas-turbine-emissions \cite{UciDataset}.}
		\label{fig:resultUci}
	\end{subfigure}
	\caption{Performances of the 4 techniques for lossless preprocessing in terms of CR on the datasets Chicago-taxi-trips-fares and UCI-gas-turbine-emissions.}
	\label{fig:results}
\end{figure}
In this section we analyze the benefits of the four proposed techniques for lossless preprocessing of floating point datasets, namely \textit{compact bins, multiply and shift, shift and separate even from odd, shift and save evenness}. In order to quantify their effectiveness, we compare the compression ratio achieved with preprocessing ($\mathrm{CR}_\mathrm{PREP}$) and without preprocessing ($\mathrm{CR}_\mathrm{NO-PREP}$), using the metric 
\begin{equation}
	\delta_{\mathrm{CR}} =\frac{\mathrm{CR}_\mathrm{PREP}- \mathrm{CR}_\mathrm{NO-PREP}} {\mathrm{CR}_\mathrm{NO-PREP}}
\end{equation}
where negative values mean that our techniques resulted in better compressed datasets. To measure the impact of metadata on the final size, we define $\mathrm{Z}$ as
\begin{equation}
	Z = \frac{\text{Metadata size in bytes}}{\text{Compressed dataset size in bytes}}.
\end{equation}
As compressor, we use Greedy-GD\cite{GD_Greedy}, since the number of common bits is particularly beneficial for its effectiveness in terms of CR. Here, $D_{M}$ is the number of desired most significant guaranteed shared mantissa bits, whereas the total number of shared mantissa bits is $S_M$, $S_E$ is for shared exponent bits and $S_\mathrm{TOT}$ for the total number shared bits. 
As input, we use the first 1000 elements of the datasets Chicago-taxi-trips-fares \cite{TaxiDataset} and UCI-gas-turbine-emissions \cite{UciDataset}, having a single dimension of non-negative floating point values:  these technique could be extended to mixed signed datasets.
In \autoref{fig:comparisonOfResults}, we summarize the best results of all 4 techniques with both datasets, while in \autoref{fig:results} we report more detailed performances.

Looking at \autoref{fig:comparisonOfResults}, we see that in all scenarios we were able to find a transformation that improved compression, since all optimal results are below zero. We also notice that the best transformation was able to achieve an improvement of $40\%$ in CR, meaning that the final size of the compressed dataset was $40\%$ smaller in size than the compressed vanilla version.

Regarding the \textit{compact bins} technique, we notice that $S_\mathrm{TOT}$ is always larger than $D_{M}$. It plateaus when all 52 mantissa bits are shared, causing $\mathrm{CR} $ to plateaus as well. We also see that $S_E$ is constant, since this technique preserves the original exponents. We see a plateau in $S_\mathrm{TOT}$ also with \textit{multiply and shift} and \textit{shift and separate even from odd}. Here, after a certain $D_{M}$ value, for every additional mantissa bit we want to be shared, we lose a shared exponent bit, making $S_\mathrm{TOT}$ constant: as a consequence, $\mathrm{CR}$ plateaus as well .

We also notice that with \textit{Compact bins} and \textit{shift and save evenness} we can impose a much higher number of common mantissa bits compared to both \textit{multiply and shift} and \textit{shift and separate even from odd}, 
since the first two require less iterations to terminate.
The trade-off is that 
$Z$ increases for larger $D$, potentially becoming larger that the size of the compressed dataset itself.

\section{Conclusions and future work}
This paper identifies key conditions for floating point addition and multiplication operations to be invertible, i.e., without introducing errors.
Using these observations, we proposed four techniques to preprocess arrays of floating point data in a lossless fashion to enhance their compression potential.
We discussed their implementations and analyzed their application on real datasets and showed their efficacy and limitations.
Our numerical results show possible compression improvements of up to $40$~\%, without introducing losses.
In the future, we plan to expand their use to mixed sign datasets and investigate their combination.

\bibliographystyle{splncs04}
\bibliography{IEEEabrv,library,IEEEtran_control}

\begin{thebibliography}{10}
\providecommand{\url}[1]{\texttt{#1}}
\providecommand{\urlprefix}{URL }
\providecommand{\doi}[1]{https://doi.org/#1}

\bibitem{IEEE754}
Ieee 754-2019 standard for floating-point arithmetic (2019)

\bibitem{dct}
Batal, I., Hauskrecht, M.: A supervised time series feature extraction
  technique using dct and dwt. In: 2009 international conference on machine
  learning and applications (2009)

\bibitem{TaxiDataset}
{City of Chicago}: Taxi trips dataset (2022),
  \url{https://tinyurl.com/4rypurjp}

\bibitem{zlib}
Jean-loup Gailly, M.A.: Zlib compressor, \url{https://www.zlib.net/}

\bibitem{UciDataset}
{H. Kaya}: Gas turbine co and nox emission dataset (2019),
  \url{https://tinyurl.com/2ubk63ra}

\bibitem{Glean}
Hurst, A., Lucani, D.E., Assent, I., Zhang, Q.: Glean:
  Generalized-deduplication-enabled approximate edge analytics. IEEE Internet
  of Things Journal  (2023)

\bibitem{GD_Greedy}
Hurst, A., Lucani, D.E., Zhang, Q.: Greedy{GD} : Enhanced generalized
  deduplication for direct analytics in {IoT} (2023),
  \url{arxiv.org/abs/2304.07240}

\bibitem{infoPreprocessing}
Klower, M., Razinger, M., Dominguez, J.J., Duben, P.D., Palmer, T.N.:
  Compressing atmospheric data into its real information content. Nature
  Computational Science  (2021)

\bibitem{FProunding}
{Mark D. Hill}: Notes on floating point arithmetic, part of course cs/ece 354
  at university of wisconsin, \url{https://tinyurl.com/mvzcyc3x}

\bibitem{FPHandbook}
Muller, J.M., Brisebarre, N., De~Dinechin, F., Jeannerod, C.P., Lefevre, V.,
  Melquiond, G., Revol, N., Stehl{\'e}, D., Torres, S., et~al.: Handbook of
  Floating-Point Arithmetic (2010)

\bibitem{additionmethod}
Taurone, F., Lucani, D.E., Feh{\'e}r, M., Zhang, Q.: Change a bit to save
  bytes: Compression for floating point time-series data. In: IEEE ICC (2023),
  \url{arxiv.org/abs/2303.04478}

\bibitem{GD_RandomAccess}
Vestergaard, R., Lucani, D.E., Zhang, Q.: A randomly accessible lossless
  compression scheme for time-series data. In: IEEE INFOCOM 2020-IEEE
  Conference on Computer Communications (2020)

\end{thebibliography}

\end{document}